\documentclass[prx,sd,amsmath,amssymb,reprint]{revtex4-1}

\usepackage{graphicx}
\usepackage{bm}
\usepackage{color}

\begin{document}


\title{Finding the stable structures of 2D hexagonal materials with Bayesian optimization:
Beyond the structural relationship with 3D crystals in weakly-bonded binary systems}

\author{Shota Ono}
\email{shota\_o@gifu-u.ac.jp}
\affiliation{Department of Electrical, Electronic and Computer Engineering, Gifu University, Gifu 501-1193, Japan}

\begin{abstract}
The graphene-graphite relationship in structural geometry is a basic principle to predict novel two-dimensional (2D) materials. Here, we demonstrate that this is not the case in binary metallic systems. We use the Bayesian optimization framework combined with the density-functional theory approach to determine the stable configuration of atomic species on a hexagonal plane. We show that the optimized structure of 2D Cu-Au exhibits the hexagonal lattice of a hexagonal ring of Cu atoms containing one Au atom, where the number of the Cu atoms is larger than that of the Au atoms in the unit cell, which is difficult to speculate from the atomic distribution of CuAu in the L1$_0$ structure. We also show that 2D Cu-$X$ with $X=$ Be, Zn, and Pd have hexagonal or elongated rings containing different atoms in the unit cell. Based on the binary Lennard-Jones model, we propose that such structures can appear for weakly-bonded systems located in between the phase-separated and strongly-bonded systems with the interatomic interaction energy between different species.
\end{abstract}

\maketitle

\section{Introduction}


A large family of two-dimensional (2D) materials has been discovered experimentally and proposed theoretically. A basic principle to find novel 2D materials is the structural relationship between 2D and 3D systems. This originates from the successful synthesis of graphene exfoliated from its bulk counterpart, graphite \cite{graphene}. This concept has been used to predict 2D materials by calculating the exfoliation energy of 2D layers from the 3D counterparts \cite{ashton}, predicting more than 600 potential 2D materials, whereas many of them are left to be synthesized. Nevalaita and Koskinen have applied similar analogy to elemental metals \cite{nevalaita}: by considering the fact that the 3D elemental metals have the close-packed structures (the fcc or hcp) as the ground state, the 2D elemental metals are expected to form the hexagonal structure. They studied the stability of 2D metals in the planar hexagonal (PHX), square, and honeycomb structures by performing the density-functional theory (DFT) calculations. They have demonstrated that among three structures the PHX structure is the lowest energy state, and that the larger the cohesive energy in the 3D phase, the larger that in the 2D phases, which establishes the 2D-3D structural relationship in elemental metals. More recently, the dynamical stability and magnetic property of elemental metals in the PHX structure have been studied using first-principles approach \cite{ono2020,ren}. 

It has not been understood whether the stable structure of 2D systems is analogous to that of 3D systems for binary systems $X$-$Y$, where $X$ and $Y$ are the metallic elements in the periodic table. This is because in the 3D bulk the $X$-$Y$ systems can form a wide variety of ordered structures depending on the composition ratio of atoms. While the author systematically studied the energetic stability of the binary 2D metals $XY$ (46 metallic elements of $X$ and $Y$) \cite{ono_satomi} and the dynamical stability of Cu$X$ \cite{ono_meta} and noble metal-based ordered alloys \cite{ono2021PRM}, only simple structures such as the buckled honeycomb structure were assumed as potential phases. 

As a model system of binary intermetallic compounds, the 3D Cu-Au has been extensively studied for many years. While the Cu and Au have the fcc structure as its ground state, the Cu-Au system forms several ordered structures depending on the mixing ratio: CuAu in the L1$_0$ structure and Cu$_3$Au and CuAu$_3$ in the L1$_2$ structure. The DFT results were consistent with such observations, while the underestimation of the formation energy by a factor of two in the Cu-Au system and the missing of the L1$_2$ CuAu$_3$ along the tie line in the phase diagram have been reported \cite{ozolins,nonlocalpbe,isaacs,ruzsinszky2019,ruzsinszky2020}. 

Recently, other ordered phases have been reported for Cu-Au systems. By using the DFT approach, Pandey {\it et al.} showed that the Cu-rich and Au-rich phases prefer the L1$_2$- and L1$_0$-derived structures, respectively \cite{pandey}. They attributed these trends to the atomic radius difference: the Au atom (the atomic radius of 1.44 \AA) is bigger than the Cu atom (1.28 \AA). Based on this fact, they interpreted that the Cu-rich structures have a large strain when the Au and Cu monolayers are stacked in a similar manner of the L1$_0$ structure. In contrast, the Au-rich structures can accommodate the smaller Cu layers, and form the L1$_0$ structure. More recently, the surface morphology of Cu-Au solid solutions and Cu$_3$Au(100) have been investigated by Liu et al. \cite{liu} and Li et al. \cite{li}, respectively.  

By performing phonon dispersion calculations, the author showed that the CuAu in the B$_h$ (WC-type) and L1$_1$ (CuPt-type) structures are dynamically stable \cite{ono_meta}. The B$_h$ and L1$_1$ structures are constructed from, respectively,  the ABAB and ABC stacking of the Cu and Au hexagonal layers that are also dynamically stable \cite{ono2020}, suggesting that the 2D layers serve as building blocks for the 3D structures. Zagler {\it et al.} reported the experimental synthesis of buckled honeycomb CuAu, where the hexagonal Au monolayer is stacked on the hexagonal Cu monolayer \cite{zagler}. This can be regarded as a 2D counterpart for the 3D metastable structures above. 

With a wide variety of 3D structures observed in the Cu-Au systems, the focus of this paper is on the 2D structures. In the present work, we explore the stable structures of 2D Cu-Au by assuming the hexagonal structure including 16 atoms in the unit cell, i.e., $4\times4$ supercell. These systems have a vast amount of configurations for the atomic species ($2^{16}$), so that the structure optimization of all configurations is a difficult issue. We use the Bayesian optimization (BO) approach to search for the lowest formation energy configuration, and find that the optimized structure exhibits the hexagonal lattice of a hexagonal ring of Cu atoms containing one Au atom, which may be difficult to speculate from the L1$_0$ structure that is the lowest formation energy structure of 3D CuAu. We also use the BO approach to search for the stable structure in the binary systems of Cu-$X$, and find that the cases of $X=$ Be, Zn, and Pd also have hexagonal or elongated rings containing different atoms in the unit cell. With the binary Lennard-Jones (BLJ) model calculations, we propose that such structures can appear for weakly-bonded systems located in between the phase-separated and strongly-bonded systems with increasing the interatomic interaction energy between different species. The present results provide interesting examples beyond the graphene-graphite relationship in structural geometry. 

In computational materials design, the BO approach has been applied to optimize various quantities such as thermal conductivity in solids \cite{seko} and nanostructures \cite{shiomi}, chemical composition of magnetic compounds \cite{fukazawa}, and foreign atom adsorption in graphene \cite{dieb} and GaN(0001) surfaces \cite{kusaba}. The present work provides another example of the BO applications. 


\section{Computational details}
\subsection{First-principles calculations}
Our calculations are based on the DFT within the generalized gradient approximation \cite{pbe}. The ultrasoft pseudopotentials was generated by using the pslibrary \cite{dalcorso}. The cutoff energy for the wavefunction and charge density was set to be 60 Ry and 600 Ry, respectively. The Monkhorst-Pack (MP) $k$ grid of $6\times 6 \times 1$ \cite{MP} and the smearing parameter of 0.02 Ry \cite{smearing} were used to construct the phase diagram (Fig.~\ref{fig_3}) of the Cu-Au system. The interlayer distance was fixed to 15 \AA \ to avoid spurious interactions in the periodic boundary condition. Spin-unpolarized calculations were performed throughout the paper. Calculations were performed using the Quantum ESPRESSO \cite{qe}. 

To study the stability of 2D $X$-$Y$, we calculated the formation energy per atom defined as
\begin{eqnarray}
 \Delta E_j(X_nY_m) = \frac{E_j (X_nY_m) - \left[ nE_0 (X) + m E_0 (Y) \right]}{n+m},
 \label{eq:form}
\end{eqnarray}
where $E_0 (X)$ and $E_0 (Y)$ are the total energy of elements $X$ and $Y$ in the PHX structure, respectively. $E_j (X_nY_m)$ is the total energy of $X_nY_m$ in the configuration $j$ with $n+m=16$. Several definitions for the formation energy have been used to study the 2D materials: the total energy of the mixed phase is subtracted by the sum of the total energy of atoms \cite{sahin}, 3D structures \cite{ono_satomi}, or 2D structures \cite{weng,ono2021PRM}. The Eq.~(\ref{eq:form}) is the same as that used in Ref.~\cite{ono2021PRM}, where it can give a negative value of the formation energy for the buckled honeycomb-structured CuAu that has been synthesized experimentally \cite{zagler}. 


When $n+m=16$, the total atomic configuration is $2^{16}=65536$, and the computational cost for investigating all configurations might be large at the DFT level. We thus remove the equivalent configurations by considering the translational, rotational, and inversion symmetry, and found 7279 configurations to be investigated. However, it is still a large number of configurations to perform DFT calculations. 

To tackle with the optimization problem above, the BO was adapted to search for the lowest formation energy structure of $X$-$Y$. We used the scikit-optimize 0.9.0 \cite{skopt} to perform the BO for the configurations of atomic species for the binary 2D systems $X$-$Y$. In the present work, each configuration was labeled by a sequence of 16 binary numbers such as 0000111100001111, where 0 and 1 indicates the atom $X$ and $Y$, respectively, and the $l$th figure from the left specifies the atomic species at the $l$th site in the unit cell on the PHX structure. The BO was done by constructing the Bayesian statistical model for the formation energy by using Gaussian process regression, where the Matern kernel was used for the Gaussian process estimator. A combination of three acquisition functions, including the expected improvement, the lower confidence bound, and the probability of improvement, were used, specifying which sample 16 binary numbers next. More than 100 steps for three runs with different random seeds were performed by assuming that the objective function is noise-free. The lattice constant $a$ and the atomic position were optimized during each step. The initial guess of the lattice constant was set to be $a(X_nY_m)=[na(X)+ma(Y)]/(n+m)$, where $a(X_nY_m)$, $a(X)$, and $a(Y)$ are the lattice constant of the $X_nY_m$, $X$, and $Y$ in the PHX structure. The magnitude of $a(X)$ and $a(Y)$ is four times larger than that of the primitive cell for the elemental metals in the hexagonal structure (see Table I in Ref.~\cite{ono2020}). 


To reduce the computational costs, the $4\times 4 \times 1$ $k$ grid was used in the configuration search based on the BO. We have confirmed that the values of $\Delta E_j($Cu$_n$Au$_m)$ are the same as those calculated by the $6\times 6 \times 1$ $k$ grid within an error of a few meV per atom, and the total energy alignment between different five models (in Fig.~\ref{fig_1} below) is not changed. 


\begin{figure*}[t]
\center
\includegraphics[scale=0.55]{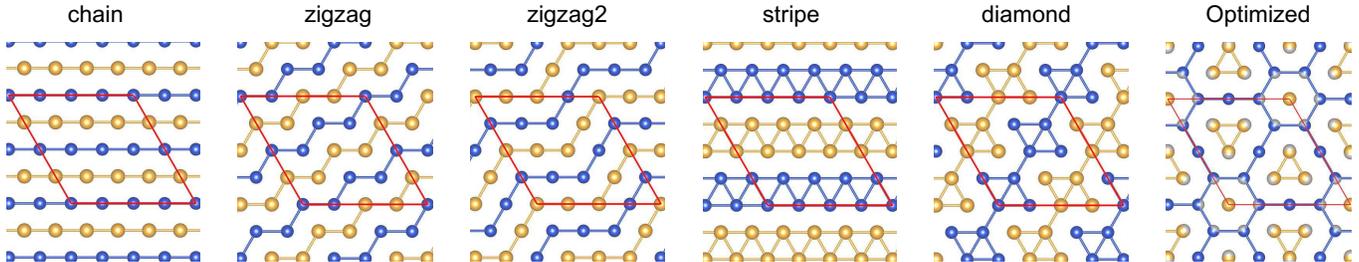}
\caption{Models of $X_8Y_8$ in the PHX structure (chain, zigzag, zigzag2, stripe, and diamond) and the optimized structure of Cu$_9$Au$_7$ obtained by the Bayesian optimization, where the blue and yellow balls indicate the Cu and Au atoms, respectively. For the relaxed geometry of the optimized structure (right), the atoms are displaced toward the gray balls. The unit cell is indicated by colored red. } \label{fig_1} 
\end{figure*}

\begin{table*}
\begin{center}
\caption{The $\Delta E_j$ (meV/atom) of the $X_nY_m$ systems for the five structures with $n=m$ (see Fig.~\ref{fig_1}) and the four structures with $n\ne m$ derived from the BO calculations. The lowest formation energy in an $X$-$Y$ is shown in bold.}
{
\begin{tabular}{lrrrrrrrrr}\hline\hline
 $X$-$Y$  \hspace{4mm} & chain \hspace{4mm} & zigzag \hspace{4mm} & zigzag2 \hspace{4mm} & stripe \hspace{4mm} & diamond \hspace{4mm} & $X_9Y_7$ \hspace{4mm} & $X_7Y_9$ \hspace{4mm} & $X_{10}Y_6$ \hspace{4mm} & $X_6Y_{10}$ \\  \hline
Cu-Ag \hspace{4mm} & $72$ \hspace{4mm} & $62$ \hspace{4mm} & $68$ \hspace{4mm} & \bm{$58$} \hspace{4mm} & \bm{$58$} \hspace{4mm} & $60$ \hspace{4mm} & $65$ \hspace{4mm} & $61$ \hspace{4mm} & $61$ \\
Cu-Be \hspace{4mm} & $-18$ \hspace{4mm} & $-25$ \hspace{4mm} & $-22$ \hspace{4mm} & $25$ \hspace{4mm} & $5$ \hspace{4mm} & $11$ \hspace{4mm} & \bm{$-29$} \hspace{4mm} & $-14$ \hspace{4mm} & $-13$ \\
Ag-Au \hspace{4mm} & \bm{$-37$} \hspace{4mm} & $-27$ \hspace{4mm} & $-32$ \hspace{4mm} & $-15$ \hspace{4mm} & $-21$ \hspace{4mm} & $-28$ \hspace{4mm} & $-25$ \hspace{4mm} & $-29$ \hspace{4mm} & $-27$ \\
Cu-Pd \hspace{4mm} & $-46$ \hspace{4mm} & $-43$ \hspace{4mm} & $-44$ \hspace{4mm} & $-19$ \hspace{4mm} & $-28$ \hspace{4mm} & $-45$ \hspace{4mm} & $-39$ \hspace{4mm} & \bm{$-51$} \hspace{4mm} & $-32$ \\
Cu-Pt \hspace{4mm} & \bm{$-65$} \hspace{4mm} & $-32$ \hspace{4mm} & $-49$ \hspace{4mm} & $0$ \hspace{4mm} & $-6$ \hspace{4mm} & $-7$ \hspace{4mm} & $-52$ \hspace{4mm} & $-23$ \hspace{4mm} & $-44$ \\
Cu-Zn \hspace{4mm} & $-69$ \hspace{4mm} & $-55$ \hspace{4mm} & $-62$ \hspace{4mm} & $-21$ \hspace{4mm} & $-41$ \hspace{4mm} & $-63$ \hspace{4mm} & $-45$ \hspace{4mm} & \bm{$-76$} \hspace{4mm} & $-48$ \\
Cu-Au \hspace{4mm} & $-66$ \hspace{4mm} & $-70$ \hspace{4mm} & $-68$ \hspace{4mm} & $-24$ \hspace{4mm} & $-46$ \hspace{4mm} & \bm{$-78$} \hspace{4mm} & $-42$ \hspace{4mm} & $-73$ \hspace{4mm} & $-54$ \\
Cu-Ga \hspace{4mm} & \bm{$-255$} \hspace{4mm} & $-253$ \hspace{4mm} & $-249$ \hspace{4mm} & $-126$ \hspace{4mm} & $-182$ \hspace{4mm} & $-251$ \hspace{4mm} & $-204$ \hspace{4mm} & $-217$ \hspace{4mm} & $-224$ \\
Cu-Al \hspace{4mm} & $-292$ \hspace{4mm} & \bm{$-304$} \hspace{4mm} & $-296$ \hspace{4mm} & $-154$ \hspace{4mm} & $-225$ \hspace{4mm} & $-282$ \hspace{4mm} & $-254$ \hspace{4mm} & $-248$ \hspace{4mm} & $-299$ \\
\hline\hline
\end{tabular}
}
\label{table_all}
\end{center}
\end{table*}

\subsection{Model calculations}
\label{sec:BLJ}
The unary LJ crystal is a simple model to describe the stability of elemental metals in the 2D and 3D structures \cite{ono_ito}, which motivates us to apply the BLJ model to interpret the stability of binary crystals $X$-$Y$. The BLJ model has been used to study the glassy materials and viscous liquids \cite{KA}, and several ordered phases have been found at zero temperature \cite{wales2001,fernandez2003,KA2018} such as coexisting phases of $X$ in the fcc structure and $XY$ in the B2 (CsCl-type) structure \cite{KA2018}. In the present work, we assumed that both types of particles have the same mass and interact via the pair potential 
\begin{eqnarray}
v_{\alpha\beta}(r_{ij}) &=& 4\varepsilon_{\alpha\beta} \left[ 
\left( \frac{\sigma_{\alpha\beta}}{r_{ij}} \right)^{12} 
- \left(\frac{\sigma_{\alpha\beta}}{r_{ij}} \right)^{6} 
\right],
\label{eq:BLJ}
\end{eqnarray}
where $\alpha$ and $\beta$ are the atomic species, $X$ or $Y$, and $r_{ij}$ is the interatomic distance between the atoms $i$ and $j$. In the present work, the BLJ crystals are assumed to have the PHX structure and contain 16 atoms in the unit cell. We used the Newton's method to optimize the lattice parameter of the hexagonal cell, and used the Broyden-Fletcher-Goldfarb-Shanno algorithm \cite{numerical_recipe} to relax the positions of 16 atoms. The Eq.~(\ref{eq:form}) is used to calculate the formation energy in the BLJ crystals. 

The pair potential in Eq.~(\ref{eq:BLJ}) has the minimum value of $-\varepsilon_{\alpha\beta}$ at $r=\sigma_{\alpha\beta}2^{1/6}$, so that the larger the $\sigma_{\alpha\alpha}$, the larger the lattice constant of the PHX structure. We regard $\varepsilon_{XX}$ and $\sigma_{XX}$ as the energy units and the length units, respectively, and chose $\varepsilon_{XX}=1$ and $\sigma_{XX}=1$. The model parameters in the BLJ model are the $\varepsilon_{XY}$, $\varepsilon_{YY}$, $\sigma_{XY}$, and $\sigma_{YY}$. The average value of the interaction length was assumed, i.e., $\sigma_{XY}=(\sigma_{XX}+\sigma_{YY})/2$. In addition, we truncated and shifted the pair potential at a cutoff distance of $2.5\sigma_{\alpha\beta}$ as in Ref.~\cite{KA}.

\section{Results and Discussion}
\subsection{Cu-Au}
\label{sec:CuAu}
To determine the lowest formation energy structure for 2D Cu-Au, we first studied the five models shown in Fig.~\ref{fig_1}: straight or zigzag chains of atoms $X$ and $Y$ are aligned alternately in the first three models, while thicker chains and clustered atoms are observed in the next two models. This is motivated by the analogy between 2D and 3D structures: when the L1$_0$ CuAu is elongated along the $c$ axis to satisfy $c/a=\sqrt{2}$, one obtains fcc CuAu, and the (111) surface consists of straight chains of Cu and Au atoms. The L1$_0$ structure is also constructed by stacking the square lattice of Cu and Au along the $c$ axis alternately. We thus expect that the chain-like structures would have the lowest formation energy structure for the Cu-Au systems. For comparison, we also studied the 2D Cu-Ag and Ag-Au, where the Cu and Ag atoms are known to be immiscible and form a separated phase, and the Ag and Au atoms can be mixed but does not form ordered phases. In this sense, the Cu$_8$Au$_8$ and Ag$_8$Au$_8$ would have the chain or zigzag structures, while the immiscible Cu-Ag systems prefer the stripe or diamond structures in Fig.~\ref{fig_1}. Table \ref{table_all} lists the values of $\Delta E_j$ for the five models, where the negative values of $\Delta E_j$ indicate that mixing different atoms is more preferable in energy. For the Cu$_8$Au$_8$ and Ag$_8$Au$_8$, the chain, zigzag, and zigzag2 structures are more stable than the stripe and diamond structures, while for the Cu$_8$Ag$_8$, an opposite tendency is observed, as expected. 


\begin{figure}
\center
\includegraphics[scale=0.35]{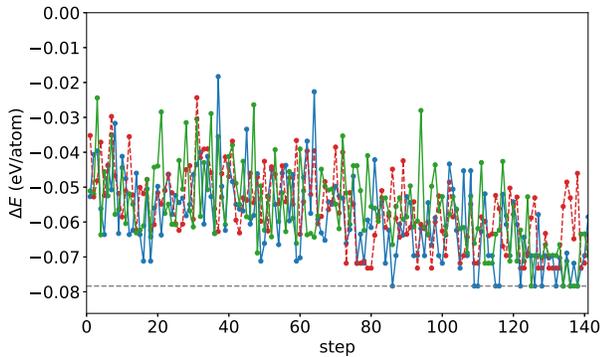}
\caption{Running the BO on the PHX Cu-Au with different three random seeds. The BO arrives at the minimum value of $\Delta E_j$ (horizontal dashed) in the 86 steps (blue solid) and 125 steps (green solid) by assuming different random seeds. The other BO (red dashed) predicts the Cu$_{10}$Zn$_6$-type structure, shown in Fig.~\ref{fig_4}. } \label{fig_2} 
\end{figure}

It is desirable to avoid the use of biased models in Fig.~\ref{fig_1}. We thus searched for the configuration showing the lowest $\Delta E_j$ by using the BO, and found the optimized configuration having the chemical formula of Cu$_9$Au$_7$ showing $\Delta E_j=-78$ meV/atom. As shown in Fig.~\ref{fig_2}, the lowest $\Delta E_j$ is obtained at about 100 steps in the BO runs. In the optimized configuration (see Fig.~\ref{fig_1}), the Cu-based hexagon ring containing one Au atom forms a large hexagonal lattice, into which upward and downward triangles of Au atoms are embedded. Among six nearest neighbor atoms, each atom has different distribution of species: an Au atom is surrounded by four or six Cu atoms, and an Cu atom is surrounded by two or three Au atoms. This is in contrast to the chain, zigzag, and zigzag2 structures, where all Au and Cu atoms are surrounded by four Cu and Au atoms, respectively, and it may be difficult to predict the optimized configuration from the analogy to the L1$_0$ structure. 

\begin{figure}[t]
\center
\includegraphics[scale=0.45]{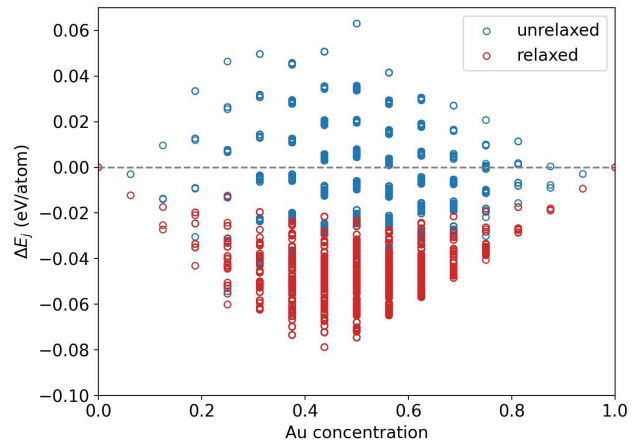}
\caption{The $\Delta E_j$ for the Cu-Au system. The values of the unrelaxed and relaxed structures are indicated by blue and red circles, respectively.  } \label{fig_3} 
\end{figure}

Figure \ref{fig_3} shows the phase diagram for the 2D Cu-Au that was constructed by 7279 configuration calculations. The minimum of the $\Delta E_j$ is observed when the Au concentration is $7/16=0.4375$, which confirms the validity of the prediction based on the BO. It should be noted that the effect of structure relaxation is strong enough to change the value of $\Delta E_j$ from positive to negative value and to change the formation energy alignment between different structures. For the unrelaxed structure, the Cu$_9$Au$_7$-type structure has $\Delta E_j=-42$ meV, while the Cu$_{10}$Zn$_6$-type structure (see Fig.~\ref{fig_4} below) shows the lowest value of $\Delta E_j=-55$ meV. The structure of the relaxed Cu$_9$Au$_7$ is depicted in Fig.~\ref{fig_1}, where the size of the Cu hexagonal ring is contracted, while that of the Au triangle is expanded. 


\begin{figure}
\center
\includegraphics[scale=0.6]{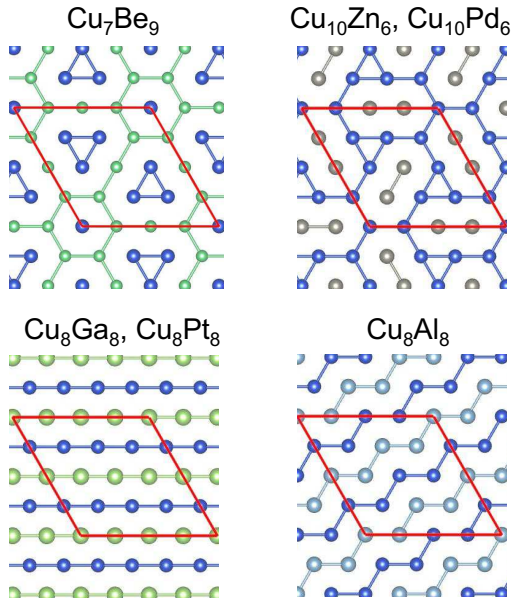}
\caption{The optimized configuration of the Cu-$X$ systems on the hexagonal plane for $X=$ Al, Be, Ga, Pd, Pt, and Zn. The blue ball indicates the Cu atom, and the solid line (red) indicates the unit cell.  } \label{fig_4} 
\end{figure}

\subsection{Cu-$X$}
Having established the relevance of the BO in the search for the lowest $\Delta E_j$ of 2D Cu-Au, we apply the same approach to explore other ordered alloys. In the present work, we focus on the Cu-$X$ systems with $X$ being a metallic element in the periodic table. We first calculated the $\Delta E_j$ in Eq.~(\ref{eq:form}) for the zigzag and stripe structures in  Fig.~\ref{fig_1}, and selected the Cu-$X$ systems having negative $\Delta E_j$. We found that the $X$ includes Al, Au, Ba, Be, Ca, Ga, Hf, In, Li, Lu, Mg, Pd, Pt, Sc, Sn, Sr, Ti, Y, and Zn. However, when $X$ is the alkali earth metals (Ca, Sr, and Ba) and the group 4 metals (Ti and Hf), the $\Delta E_j$ of the stripe structure is lower than that of the zigzag structure. This implies that the energy gain is small when the Cu and $X$ atoms are alloyed, so that we exclude these $X$ below. 

We next applied the BO approach to search for the optimized mixing between the Cu and $X$ atoms in the $4\times 4$ sites on the hexagonal plane. When the atomic size mismatch is large between the Cu and $X$ atoms, many steps will be needed in the structure relaxation within the DFT. In the present study, we studied the atom $X$ satisfying the condition that the relative error of the lattice constant between the 2D Cu ($2.428\times 4$ \AA) and 2D $X$ in the PHX structure is less than 10 percent. We thus studied the cases of $X=$ Al, Be, Ga, Pd, Pt, and Zn, in addition to $X=$ Au. 

Figure \ref{fig_4} shows the optimized configurations of the Cu-$X$ on the hexagonal plane for the $X$ selected above. For the Cu-Be system, the Be-rich structure, Cu$_7$Be$_9$, has the lowest value of $\Delta E_j$, and is isostructural with the Cu$_9$Au$_7$ found by the BO, with Cu on the Au site and Be on the Cu site. This is due to the similar value of the relative difference of the lattice constant for the elemental metals in the PHX structure: The value of $\Delta a = \vert a_{\rm PHX}({\rm Cu}) - a_{\rm PHX}(X)\vert /a_{\rm PHX}({\rm Cu})$ is estimated to be 13.2 \% for $X=$ Au and 12.4 \% for $X=$ Be, where $a_{\rm PHX}({\rm Au})=2.748\times 4$ \AA \ and $a_{\rm PHX}({\rm Be})=2.126\times 4$ \AA \ were assumed. For the Cu-Zn and Cu-Pd systems, there are elongated hexagons containing the Zn and Pd dimers, where the three elongated hexagons share a triangle of the Cu atoms. As mentioned, this structure has the lowest value of $\Delta E_j$ for the unrelaxed Cu-Au system. In the running of the BO for the Cu-Ga, Cu-Pt, and Cu-Al systems, the chain or zigzag structures have the lowest value of $\Delta E_j$. The values of $\Delta E_j(X_nY_m)$ for several $j$s are listed in Table \ref{table_all}. It should be noted that weakly-bonded systems ($-100 \lesssim \Delta E_j < 0$ meV) tend to show the formula of $X_nY_m$ with $n\ne m$. This assumption will be rationalized below by using the BLJ model.  

Nepal {\it et al}. have proposed that the $X$-$Y$ system is weakly-bonded when $X$ and $Y$ have completely filled $d$-bands, by assuming the B2 structure and performing DFT calculations \cite{ruzsinszky2019,ruzsinszky2020}. Such combinations of $X$ and $Y$ include the Cu-Au, Cu-Zn and Cu-Pd systems, which are consistent with the present findings. In addition, they also proposed that the Ag-Zn, Ag-Cd, Au-Cd, and Au-Zn are weakly-bonded systems.

In respect to the composition ratio, the analogy between 2D and 3D structures cannot hold for Cu-Zn, and Cu-Pd systems. By referring to the Materials Project database \cite{materialsproject}, the Zn-rich phase of Cu$_{17}$Zn$_{35}$ and the CuPd in the B2 structure have the lowest formation energy for Cu-Zn and Cu-Pd systems, respectively. On the other hand, the Be-rich phase was found to be the most stable structure in both 2D and 3D Cu-Be systems, where the latter is identified as CuBe$_2$ in the C15 (Laves) structure. 

\begin{figure*}
\center
\includegraphics[scale=0.6]{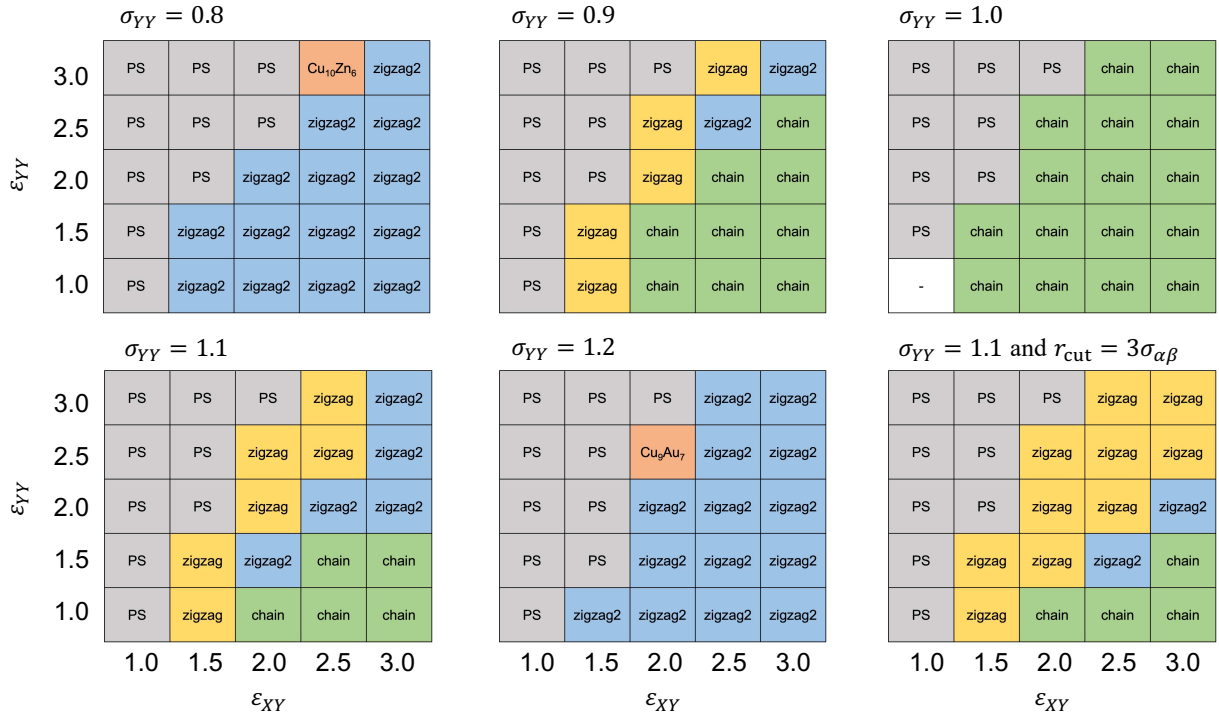}
\caption{The lowest energy structures for the BLJ crystal in the PHX structure when the pair potential parameters of $\varepsilon_{XY}$, $\varepsilon_{YY}$, and $\sigma_{YY}$ are tuned. The cutoff radii for the interatomic interaction is set to be $r_{\rm cut}=2.5\sigma_{\alpha\beta}$ except for the lower right. ``PS'' stands for the phase separation, and the hyphen in $(\varepsilon_{XY}, \varepsilon_{YY}, \sigma_{YY})=(1,1,1)$ indicates the unary LJ crystal. } \label{fig_5} 
\end{figure*}

\begin{figure}
\center
\includegraphics[scale=0.45]{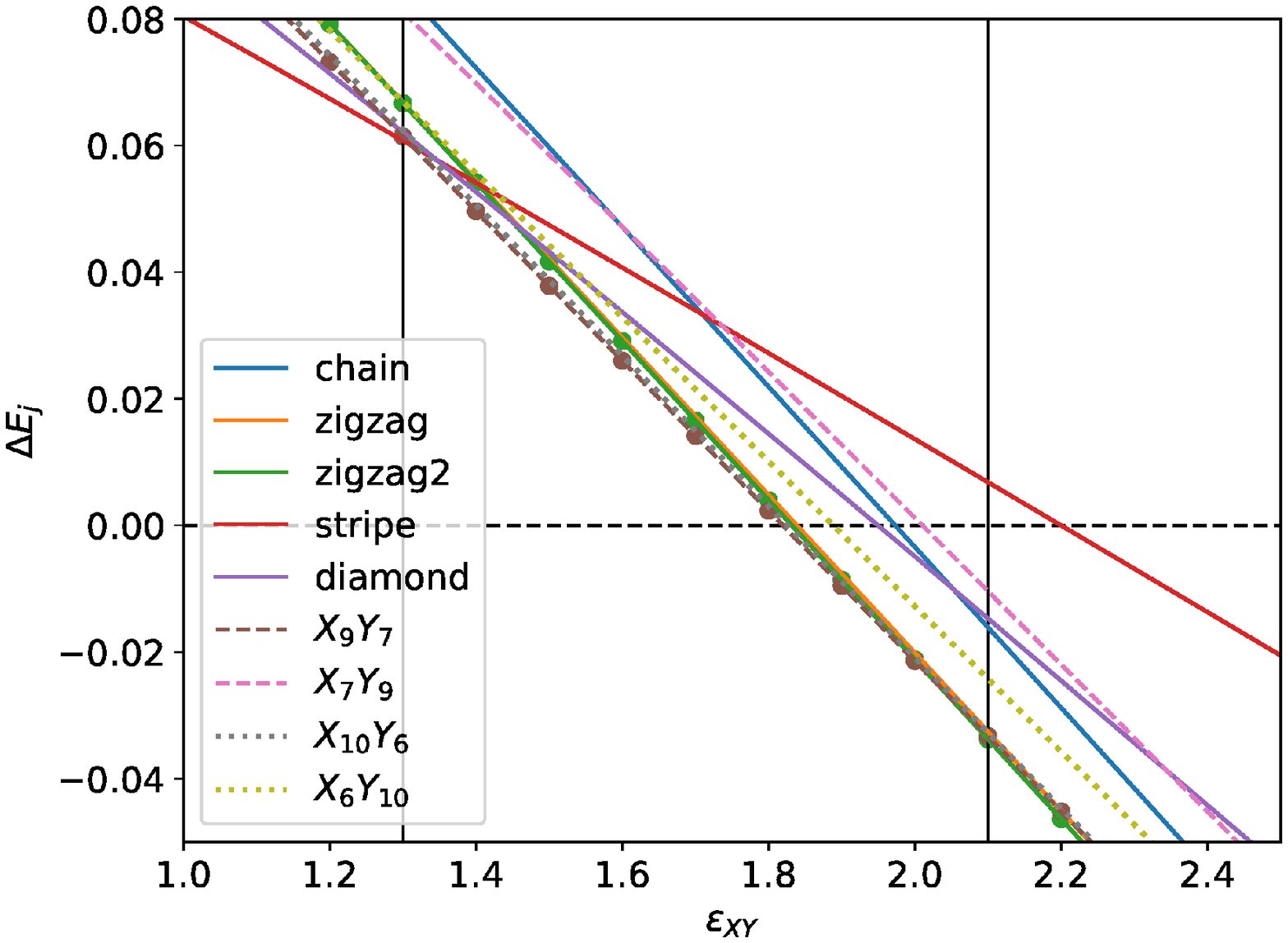}
\caption{The $\varepsilon_{XY}$-dependence of $\Delta E_j$ for the BLJ crystals for nine structures, when $\varepsilon_{YY}=2.5$ and $\sigma_{YY}=1.2$ (see the lower middle panel in Fig.~\ref{fig_5}). The calculated data for the $X_9Y_7$ and zigzag2 structures are also shown by circles. The lowest energy structure changes from the stripe to $X_9Y_7$ to zigzag2 across the vertical lines. The horizontal dashed indicates $\Delta E_j=0$. } \label{fig_6} 
\end{figure}

\subsection{Binary Lennard-Jones}
Without prior knowledge on the configurations predicted by the BO, it may be difficult to find the Cu$_9$Au$_7$-type and Cu$_{10}$Zn$_6$-type structures (in Fig.~\ref{fig_4}) because the condition $n\ne m$ in Cu$_nX_m$ holds. To understand when the structures with $n\ne m$ are more stable than those with $n=m$, we calculated the $\Delta E_j$ for the BLJ crystals, as described in Sec.~\ref{sec:BLJ}. We studied the $X_8Y_8$ in the first five structures as shown in Fig.~\ref{fig_1}, the $X_9Y_7$ (Cu$_9$Au$_7$-type) structure (see Fig.~\ref{fig_1}), the $X_{10}Y_6$ (Cu$_{10}$Zn$_6$-type) structure (see Fig.~\ref{fig_4}), and the isostructural configurations (i.e., $X_7Y_9$ and $X_6Y_{10}$). Figure \ref{fig_5} shows the lowest formation energy structures as a function of $\varepsilon_{XY}$ and $\varepsilon_{YY}$ for several $\sigma_{YY}$. When $\varepsilon_{YY}\gg\varepsilon_{XY}$, the phase separation (i.e., $\Delta E_j >0$) is preferred because the interatomic interaction between the $X$ and $Y$ atoms is not strong enough to alloy with different species. When $\varepsilon_{XY}$ is increased, the $X$ and $Y$ atoms are mixed, forming an ordered structure that depends on the parameters of $(\sigma_{YY},\varepsilon_{XY},\varepsilon_{YY})$: When $\sigma_{YY}=1.0$, the chain structure is the lowest formation energy structure for $\varepsilon_{YY}<\varepsilon_{XY}$; when $\sigma_{YY}=0.9$ and 1.0, the zigzag structure appears around the boundary $\varepsilon_{YY}\simeq \varepsilon_{XY}$. In addition, another zigzag structure with a longer period (zigzag2) appears for large $\varepsilon_{XY}$; and when $\sigma_{YY}=0.8$ and 1.2, such a long-period zigzag structure becomes the lowest formation energy phase for $\varepsilon_{YY}<\varepsilon_{XY}$. We also studied the cases with the cutoff radius of $3.0 \sigma_{\alpha\beta}$. However, the phase diagrams are basically the same as those calculated by using $2.5 \sigma_{\alpha\beta}$, while the zigzag structure overcomes the zigzag2 structure for large $\varepsilon_{XY}$. 

It is noteworthy that the Cu$_{10}$Zn$_6$-type and the Cu$_9$Au$_7$-type structures can be found when $(\sigma_{YY},\varepsilon_{XY},\varepsilon_{YY})=(0.8, 2.5, 3.0)$ and $(\sigma_{YY},\varepsilon_{XY},\varepsilon_{YY})=(1.2, 2.0, 2.5)$, respectively. Although the inequality $\varepsilon_{YY}>\varepsilon_{XY}$ holds, moderately large value of $\varepsilon_{XY}$ barely prevents the $X$-$Y$ system from the phase separation. To study the impact of $\varepsilon_{XY}$ on the energy alignment, we plotted the values of $\Delta E_j$ for several structures in Fig.~\ref{fig_6} by assuming $(\varepsilon_{YY},\sigma_{YY})=(2.5,1.2)$. As $\varepsilon_{XY}$ increases, $\Delta E_j$ decreases, and becomes negative value around $\varepsilon_{XY}=1.8\sim 2.2$ depending on the structure. This shows that in the limit of small and large $\varepsilon_{XY}$ the $X$-$Y$ system can be regarded as an immiscible and a strongly-bonded system, respectively, and for moderate $\varepsilon_{XY}$ the $X$-$Y$ system is a weakly-bonded system. Among nine models, the lowest formation energy structure changes from the stripe to $X_9Y_7$ (at $\varepsilon_{XY}=1.3$) to zigzag2 (at $\varepsilon_{XY}=2.1$) structures with $\varepsilon_{XY}$. The difference of $\Delta E_j$ between the $X_9Y_7$ and $X_{10}Y_6$ structures are small (less than 0.001 for $\varepsilon_{XY}\ge 1.4$). In this respect, the weakly-bonded systems prefer the Cu$_9$Au$_7$-type and the Cu$_{10}$Zn$_6$-type structures. Furthermore, the BLJ model calculations are consistent with the small values of $\Delta E_j$ in the Cu-$X$ with $X=$ Au, Be, Pd, and Zn (larger than $-100$ meV/atom listed in Table \ref{table_all}). 

 

\section{Conclusion}
We have demonstrated that the combined use of DFT and BO approach enables us to predict the lowest formation energy structure of the Cu-Au system in the PHX structure with 16 atomic sites. The stable structure has the chemical formula of Cu$_9$Au$_7$, which is a counterintuitive distribution pattern of Cu and Au atoms because no structural similarities with the L1$_0$ CuAu can be found. We also explored the stable structures of Cu$_nX_m$ with $n+m=16$ and $X=$ Be, Al, Zn, Ga, Pd, and Pt, in which the lattice constant of $X$ in the PHX structure is similar to that of Cu. The lowest formation energy structure satisfies the condition $n\ne m$ when $X=$ Be, Pd, and Zn. The BLJ model calculations suggest that such structures can appear for weakly-bonded systems located in between the phase-separated and strongly-bonded systems. It is desirable to explore the stable structures for multilayered systems, leading to a fundamental understanding of the structural similarity between 2D and 3D systems. We expect that many interesting structures will be found when the boundary condition (i.e., the PHX structure containing 16 atoms) is removed, or when other $X$-$Y$ systems are considered. 


The present approach can be extended to investigate the stable spin configurations of magnetic materials in the PHX structure, which will enable us to find the lowest energy anti-ferromagnetic phase that overcomes the stability of stripe order in 2D elemental metals \cite{ren}. It is interesting to determine the noncollinear spin configuration of 2D Mn (i.e., including spin-orbit coupling) because the 3D Mn has a cubic cell containing 58 atoms, showing a complex anti-ferromagnetic ordering in the ground state \cite{Mn2020}.

In the present work, we assumed the planar structure, so that the 2D Cu-$X$ might be unstable to the out-of-plane vibrations called the flexural phonon modes. We expect that these systems may be stabilized by using monolayer and/or bilayer graphene: For example, one-forth of the lattice constant of the optimized Cu$_9$Au$_7$ ($\simeq 2.58$ \AA) is close to the lattice constant of graphene (2.46 \AA), so that the Cu$_9$Au$_7$ can be realized on the graphene substrate, as in the experimental synthesis of buckled honeycomb CuAu (2.68 \AA \ \cite{zagler}). Another possibility is the use of the 2D nanospace realized by the gap of the bilayer graphene \cite{kanetani,ago}, where the effect of the van der Waals interaction forces between the 2D systems and the graphene may be important to stabilize the out-of-plane vibrations. 

\begin{acknowledgments}
This work was supported by JSPS KAKENHI (Grant No. JP21K04628). The computation was carried out using the facilities of the Supercomputer Center, the Institute for Solid State Physics, the University of Tokyo, and using the supercomputer ``Flow'' at Information Technology Center, Nagoya University.
\end{acknowledgments}





\end{document}